\documentclass[12pt,preprint]{aastex}

\newcommand{\kms}{km~s$^{-1}$}

\newcommand{\etal}{et~al.~}

\begin{document}
\title{Kinematics of NGC 2264: signs of cluster formation \altaffilmark{1}}

\author{G\'{a}bor F\H{u}r\'{e}sz\altaffilmark{2,3,4},
Lee W. Hartmann\altaffilmark{5},
Andrew H. Szentgyorgyi\altaffilmark{2},
Naomi A. Ridge\altaffilmark{2},
Luisa Rebull\altaffilmark{6},
John Stauffer\altaffilmark{6},
David W. Latham\altaffilmark{2},
Maureen A. Conroy \altaffilmark{2},
Daniel G. Fabricant\altaffilmark{2},
John Roll\altaffilmark{2}
}

\altaffiltext{1}{Observations reported here were obtained at the MMT Observatory, a 
joint facility of the Smithsonian Institution and the University of Arizona}
\altaffiltext{2}{Center for Astrophysics, 60 Garden Street, Cambridge, MA 02138}
\altaffiltext{3}{most of this work was done while GF was a research fellow of the Konkoly Observatory of the Hungarian Academy of Sciences,
P.O. Box 67, H--1525 Budapest, Hungary}
\altaffiltext{4}{University of Szeged, Department of Experimental Physics, 
Dom ter 9, H--6723 Szeged, Hungary}
\altaffiltext{5}{Dept. Astronomy, University of Michigan, 500 Church St., 830 Dennison Building, Ann Arbor, MI 48109}
\altaffiltext{6}{SIRTF Science Center, Mail Stop 220--06, California Institute
of Technology}

\begin{abstract}
We present results from 1078 high resolution spectra of 990 stars in the 
young open cluster 
NGC 2264, obtained with the Hectochelle multiobject echelle spectrograph on the 6.5m 
MMT. We confirm 471 stars as members, based on their radial velocity and/or
$H\alpha$ emission. The radial velocity distribution of cluster members
is non-Gaussian with a dispersion of $\sigma \approx 3.5 $\kms.  
We find a substantial north-south velocity gradient and spatially coherent structure in
the radial velocity distribution, similar to that
seen in the molecular gas in the region. Our results suggest that there
are at least three distinguishable subclusters in NGC 2264, correlated with similar
structure seen in $^{13}$CO emission, which is likely to be a remnant of initial 
structure in this very young cluster.  We propose that this substructure is
the result of gravitational amplification of initial inhomogeneities during
overall collapse to a filamentary distribution of gas and stars, as found
in simulations by \citet{bh04}.
\end{abstract}


\section{Introduction}

Studies of young galactic clusters associated with molecular clouds can provide insight
into the early history and formation of these objects.  One of the most famous
and best studied young cluster is NGC 2264. 
Many investigations of this region have been undertaken,
building on the early work of Herbig (1954).  A number of studies suggest low
extinction toward NGC 2264 ($E(B-V)=0.082$ by Walker (1956);
$E(B-V)=0.071 \pm 0.033$ by Sung et al. (1997); $E(B-V)=0.066 \pm 0.034$ by Park et al. (2000)),
although the large amount of molecular gas in the region (\S 2.4) suggests that a substantial
number of sources embedded in dust remain to be found.
The optical nebular emission is less intense 
than in the Orion Nebula Cluster (ONC), so contamination of stellar
spectra from nebular emission is less of a problem in NGC 2264.
There is a dark cloud immediately behind the 
cluster which obscures most of the background stars, thus reducing the
number of potential impostor members \citep{herb54}. 
Thus, despite the fact that NGC 2264 is more distant
than the ONC and other nearby regions of star formation 
( 800 pc by Walker 1956; 800 pc by Sagar \& Joshi 1983; $950\pm75$ pc by P\'erez et al. 1987;
$760\pm40$ pc by Park et al. 2000) the low extinction and weak nebular emission make NGC 2264
a favorable and important target for star and cluster formation studies. 

Age estimates of NGC 2264 have been reduced as the quality of NGC 2264 observations
have improved. Early 
estimates suggested ages 3-5 Myr \citep{hari76,park00,palla99},
while \citet{flac00} estimate an age closer to
1-2 Myr, and \citet{rami04} found that most of the X-ray stars were younger
than 3 Myr.  A major reason for the discrepancy between different estimates is
the differing weight placed on specific mass ranges. 
As pointed out by \citet{hart03},
ages inferred from $1 - 2$ M$_{\bigodot}$ stars 
in young associations tend to be older
than those of the lower-mass stars.  \citet{hart03} argued that the ages of
these slightly higher-mass stars are systematically overestimated by neglecting
``birthline'' effects (e.g., pre-main sequence contraction from finite starting
radii); this leads us to favor the lower ages of \citet{flac00}. and \citet{rami04}
based on the lower mass stars, which are more consistent with the large amount
of molecular gas associated with the cluster.

The spatial distribution of the stars in NGC 2264 is extended 
along the north-south direction,
paralelling the extension of  the molecular gas in the region \citep{ridg03}.  
This feature, consistent with the general youth of the cluster, suggests that
NGC 2264 may still exhibit the morphology of the gas cloud from which the
protostars coalesced. 
The early evolution of stellar systems may be strongly affected
by the density and dynamics of the environments in which they form. 
Numerical simulations \citep{scal02} show that the lifetime of any primordial
substructure is short, a few million years or less.  Radial velocity studies
may be able to provide clues to cluster formation for very young regions by revealing coherent patterns in the motion of sub-regions.
 
In this paper we report the initial results of our continuing 
radial velocity (RV) survey of NGC 2264, in the search for
kinematic signs of primordial structures in young stellar clusters. 
In \S 2 we describe the observations and data reduction; in \S 3 we compare the
radial velocity distribution of the stellar and gaseous component, the latter as traced
by $^{13}$CO measurements.  Finally, we discuss possible implications of our results for cluster formation scenarios.

\section{Observations and data reduction}

\subsection{Target selection}

We drew our target sample from several sources, starting with surveys of H$\alpha$
emission-line objects (Herbig 1954; Reipurth et al. 2004 -- R04 hereafter), 
narrow-band photometry \citep{sung97,sung04,park00}, and
samples of young objects selected based upon their IR and UV excess 
\citep{lada93,rebu02}.
All of these  surveys have a bias toward finding (accreting) 
Classical T Tauri stars (CTTS) and against finding pre-main 
sequence (PMS) stars with weak or no disks.
Therefore, by supplementing these samples with a large
catalog of X-ray sources \citep{flac00,rami04},
also sensitive to weak-emission (non-accreting) T Tauri stars (WTTS),
the bias towards accreting stars can be corrected.
These surveys, together with proper motion 
\citep{vasi65} and variability \citep{lamm04} surveys 
provide a very good input catalog of potential cluster members.

The selected targets come from three of the catalogs above: a substantial fraction,
1389 stars, from \citet{rami04} X-ray sources; 408 H$\alpha$ emission stars from R04; 
and 100 stars from \citet{park00}. As some of the stars appeared in more than one of
these three cataloges, the final target list contained 1563 stars.

\subsection{Spectroscopic observations}

The stellar spectra were obtained with  Hectochelle \citep{sain98}, 
a fiber-fed, bench mounted echelle spectrograph
that operates at the post-conversion MMT. The MMT is a 6.5m Cassegrain
telescope and the Hectochelle is designed to operate with the MMT 
in  its f/5, wide field mode \citep{fabr04}. The 
robot positioner places all 300 fibers buttons for observation in 
five minutes with an accuracy of 25$\mu$m or better. The fiber diameter
is 250$\mu$m, or 1.5\arcsec~ at the MMT f/5 plate scale. The resolution of the
Hectochelle is $\sim$34,000 and the the peak efficiency is 7\% of the light
striking the primary mirror.
Because the Hectochelle and Hectospec, a moderate dispersion fiber fed
spectrograph,  both use the same fiber feed,
but have different magnifications, only 240 fiber images fall on the 
Hectochelle focal plane format, while Hectospec exploits all 
300 available fibers.
  
The preferred order for radial velocities with Hectochelle is
centered at 5225{\AA} and includes the Mg triplet.
Unfortunately, a failure of the original coatings (now replaced) on the camera and 
collimator mirrors of Hectochelle
reduced the reflectivity in this spectral region by a factor of up to 6 
per mirror. Also, our targets have brighter continuua at H$\alpha$ than
at 5225{\AA}, helping to compensate for fewer lines.
Therefore we decided to use the 190{\AA} wide order centered at H$\alpha$, 
which was relatively unaffected by the coating failure.
Although this order has a smaller number of lines than the 5225{\AA}
order, we found that by masking out the $H\alpha$ line when 
executing the cross-correlation
we can get velocities with an error of $1-2$ \kms. This 
velocity resolution is good enough for our
kinematic study (see, e.g., Sicilia-Aguilar et al. 2005). 
While this order is not optimal for radial
velocity measurement, observations of the 
$H\alpha$ profiles allows us to unambiguously identify and distinguish CTTS and WTTS stars.\

Our first observations in NGC 2264 were taken in March 2004. These data were
obtained during the commissioning phase of the instrument, and we ended up observing 155 targets     
in one fiber configuration (see Table \ref{table:obs})
centered north of the Cone nebula.
The preliminary results derived from this first data set showed interesting results
so we decided to observe NGC 2264 again.
We revisited the cluster selecting targets from an expanded list
including objects from the cataloge of R04 and the full cataloge of \citet{rami04}.\ 

Even though the spectrograph mirrors were re-coated during the summer of 2004 and 
we could have used the 5225{\AA} order, we found that the accuracy of radial velocities 
obtained in the H$\alpha$ order to be adequate and the extra information in the
H$\alpha$ line profiles
to be scientifically  useful so we continued our observations in this order.
The second observing run was carried out in December, 2004. During the nights
of 2 Dec. and 3 Dec. we observed 923 targets in four fields 
(see Table \ref{table:obs}), including 88 targets observed in 
March which were thought to be members based on the March RV data.\

We have found, in observations of regions with strong, spatially structured nebular
emission, e.g. the Orion Nebula Cluster  \citep{auro04}, that it is necessary to offset the
telescope 10\arcsec~ without moving the fiber buttons to obtain a good
measurement of nebular emission contaminating the stellar spectra.
In March we followed this procedure.
We found that in NGC 2264  this background subtraction is only important near
the H$\alpha$ and [N II] lines; it is unimportant for radial velocity measurements,
which depend only upon the absorption line spectra in between the emission features.
Therefore in December we opted to eliminated the offset observations
so that we might spend more time collecting stellar spectra.\

\subsection{Radial velocities}

To establish the radial velocity zero point for our
spectra and to provide a variety of well-observed spectra that could
serve as templates in the cross-correlation analysis, we obtained spectra of
10 relatively bright stars in a field located in Selected Area 57 \citep{pick18},
near the north Galactic Pole.  
The radial velocities of these stars
have been monitored for many years using the CfA Digital Speedometers
\citep{Lath92} without showing any signs of variation, and their
absolute velocities have all been measured with a precision better
than 0.2 \kms~ following the procedures outlined by \citet{stef99}.
These 10 stars were included in the sample discussed by
\citet{Latetal91}, but two recent technical developments have made it possible
to improve measurements of their  mean velocities.
First, several additional observations had been
obtained over the intervening years, thus doubling the typical time
spanned by the radial velocities measurements.  Second, the spectra were
re-analyzed using our latest library of synthetic spectral templates calculated
by Jon Morse for a grid of \citet{kuru79} model atmospheres for a range
of effective temperatures, surface gravities and rotational velocities. Solar
abundance is assumed in all cases.
We ran grids of correlations between these templates and observed
spectra using the \textit{rvsao} package
\citep{kurt98} running inside the IRAF \footnotetext[1]{IRAF (Image
Reduction and Analysis Facility) is distributed by the National
Optical Astronomy Observatories, which are operated by the Association
of Universities for Research in Astronomy, Inc., under contract with
the National Science Foundation.}environment.  For the final velocity
determinations, we adopted the template parameters that gave the
strongest correlation peak averaged over all
the observed spectra (see Table \ref{table:sa57}; and Stefanik et al., in prep.).

To calibrate the wavelength scale of our spectra, we used
Thorium-Argon lamp exposures for both the NGC 2264 targets and the SA57
secondary standards. We adopted the observed
spectrum of W23870 \citep{Latetal91} in SA57 as the template for the
velocity determinations of all our NGC 2264 targets in March, because this K
dwarf gave the best results using the cross-correlation technique. 
However, for the final reduction of the entire  data set we made 
use of all the template stars in the SA57 field, in order to 
obtain the most accurate radial velocities.
For each of the observed NGC 2264 targets we selected the 
``best matching template'' which yielded the correlation 
peak with the highest signal-to-noise value. 
If the RV values from all the other 9 templates were comparable 
to the radial velocity measured with this best matching template, 
we combined all the 10 measurements of a 
given star and used the median value for the further analysis.\

\begin{figure}
\epsscale{1.0}
\plotone{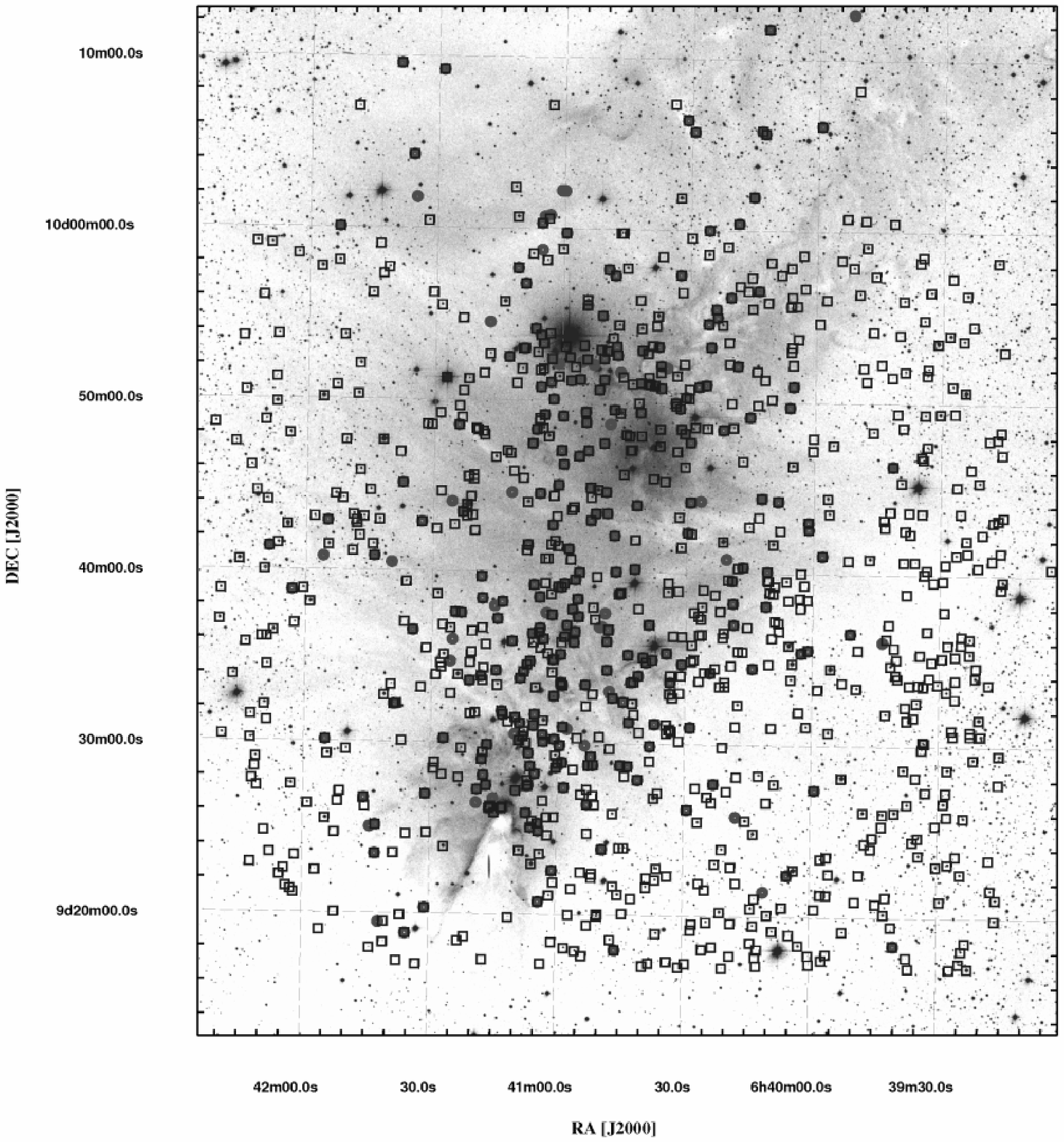}
\caption{Our targets in the field of NGC 2264 (open diamonds),
superimposed on a DSS POSS1 red plate image of the cluster.
$H\alpha$ emission stars (filled circles) from R04 are also
shown to guide the eye where the concentration of young stars
expected to be larger. Most of these emission stars were observed
by us, but as can be seen our targtes were more evenly distributed,
covering the outer regions of the cluster as well.
\label{fig:optical}}
\end{figure}

Using this method of template selection, we obtained radial velocities 
with errors less than $\pm$ 1.5 \kms~ for 344 targets,
as shown in Fig. \ref{fig:optical}. In this figure we also plot the
H$\alpha${} emission stars of R04 objective prism survey
to guide the eye where the larger concentrations of cluster members are
expected.\
 
In order to ensure consistency between the March and December data
we observed 88 stars in both runs. For 31 of these stars yielding the
highest correlation peaks 
the mean of the velocity differences is $RV_{Mar}-RV_{Dec}=0.3$ \kms~ with
an RMS of 0.8 \kms.\

\subsection{$^{13}$CO observations}

Observations of the $^{13}$CO transition of CO (110.201\,GHz) were
obtained at the Five College Radio Astronomy Observatory (FCRAO) 14m
telescope located in New Salem, Massachusetts during 2002 September, as
part of a larger survey of nearby cluster forming regions \citep{ridg03}.

Using the SEQUOIA 32-element focal-plane array and On-the-Fly (OTF)
mapping technique eight $15'\times 15'$ submaps were obtained over 3
nights, in order to build up the final $30'\times 60'$ map.  Submaps
were obtained by scanning in RA, and an
``off'' source reference scan was obtained after every two rows.  The
off-position was checked to be free of emission by performing a single
position-switched observation with an additional 30$'$ offset. Each
submap was repeated twice to increase the sensitivity.

The narrow band digital correlator provided a total bandwidth of
25\,MHz over 1024 channels, yielding an effective velocity resolution
of 0.07\,kms$^{-1}$. System temperatures at the observing frequency
were between 200 and 300\,K (single sideband). Calibration was via the
chopper-wheel technique (Kutner \& Ulich 1981), yielding spectra with
units of {{T$_A$}$^*$}. The final combined map has an average rms
sensitivity of ~0.2\,K/channel. A full discussion of the data analysis
procedure and the integrated intensity image was presented in \citet{ridg03}.

\section{Results}

\subsection{Distribution of radial velocities \label{results_1}}

\begin{figure}
\includegraphics[angle=270,scale=0.6]{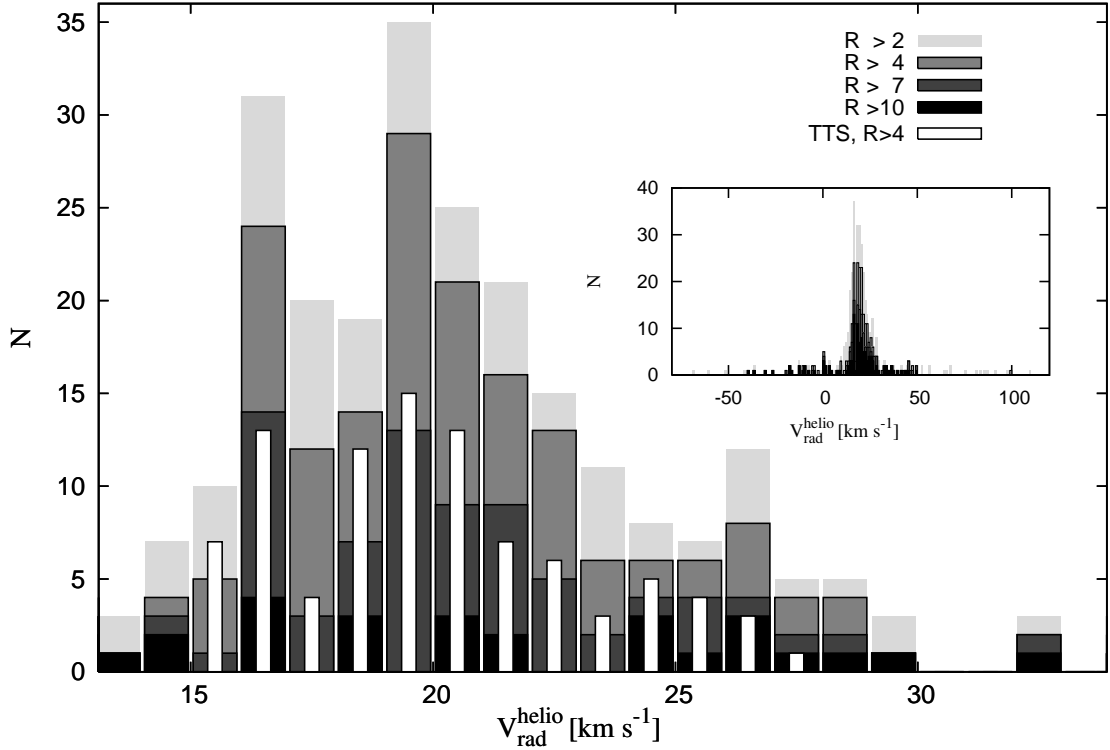}
\caption{Radial velocity distribution for 344 of the observed stars which
were giving consistent velocities independent of the RV template. The
distribution is unusually wide and non-gaussian.
The insert is an expanded velocity range to show how clearly the cluster
members form a peak in the velocity space.
There are four selections of stars plotted according to the signal-to-noise (R)
values of the correlation, showing that the large dispersion is real and not
coming from the uncertainties of RV measurements (see text for details).
The white impulses show a selection of stars with $H\alpha$ emission
presented in their spectra. The similarity between the distribution of these 
and non emitting stars suggest that even a selection is biased towards TTS
it can be representative of the entire cluster in kinematical studies.
\label{fig:hist}}
\end{figure}

The radial velocity distribution of the 344 stars with accurate RV values is
shown on Fig. \ref{fig:hist}. The cluster members clearly 
form a peak centered around $V^{helio}_{rad} = 22 $ \kms, but the
distribution is unusually wide and not symmetrical. 
We also report the distribution on an expanded velocity
scale to show how clearly the cluster members form a peak in radial
velocity space. \

We plot the RV distribution for four selections of stars
according to the signal-to-noise values \citep{tonr79} of the correlation, in Fig. \ref{fig:hist}. This R value is inversely proportional 
to the error of velocity measurement. As described in \citet{kurt98} 
the errors consist of a systematic and a statistical component,
the  systematic could be determined for a given instrument by comparing radial 
velocities measured on large number (several hundred) of objects at two epochs.
A very good estimate for the random velocity error is $\sigma=(3/8)\ast[w/(R+1)]$ \citep{kurt98},
where \textit{w} is the FWHM of the correlation peak and \textit{R}
is the signal-to-noise ratio. 

This estimate agrees well with our result in NGC 2264 based on 31 stars, observed both in
March and December with an R value larger than 6 (see \S 2.3).
Also, based on results with other instruments,  
the estimate for the error mentioned above is likely within $\pm 30$\% of the real error value for a Hectochelle measurement. 
We therefore adopt a velocity error of
$\sigma = 6/(R+1) \, {\rm km\, s^{-1}}$; the error is larger for
stars with rotational velocities greater than about 10~\kms.
In principle, this means that slowly-rotating stars with
correlation peaks of $R=2$ can yield a velocity error of 2 \kms. 

Of course, in our spectral order of velocity width $\sim 8000$ \kms,
we can expect to find, just due to chance, roughly one $3 \sigma$ peak 
(with an FWHM of $\sim 10$~\kms~ or an R value of $\sim 3$)
in the cross-correlation function, somewhere within the interval.
If we restrict our attention to an interval of $\sim 100$~\kms~ near
the median cluster velocity, then we would expect to have a probability
of about 10\% of finding a $R \sim 2$ peak of width $\sim 10$~\kms.
The $R > 2$ selection is included in Fig. \ref{fig:hist} because,
even with the larger uncertainty in velocity, the distribution of these stars  follow the overall shape of the histogram fairly well.
As a reminder that large errors are associated with
this group, we plotted this group as light gray columns
without a border.\
There is a fifth sample of stars displayed on the left panel of Fig. \ref{fig:hist}
as narrow, white bars: stars with $H\alpha$~ emission detected in
their spectra. Comparing this sample to the others we can say that the radial velocity
ditribution of TTS is the same as for the rest of our targets. Also, as the 
distribution is same independently of the R value of the correlation, the observed
large dispersion is not due to uncertainties in the RV measurement. \\

The RV histogram of Fig. \ref{fig:hist} is far from Gaussian. To estimate the velocity dispersion we used the definition $\sigma^{2}\equiv\sum{(x_{i}-x)}^{2}/(N-1)$. Assuming lower and upper limits for membership as 13 \kms and 30 \kms, the estimated one-dimensional RV dispersion is $\sigma \approx 3.5$~\kms.
This is much larger than in the ONC, for example, which has a dispersion of 2.3 \kms~
\citep{jone88}, and $\sim 1.8$~\kms~ for the brighter stars (see also Sicilia-Aguilar
et al. 2005). The skewness of the velocity distribution suggests that the cluster is not dynamically relaxed, consistent with our finding of spatially-coherent motions (\S 3.3).\

\subsection{$H\alpha$~ emission and binary stars}

Recording spectra at the $H\alpha$ order has the advantage of not just measuring RV, but also identifying T Tauri stars and distinguishing between the CTTS and WTTS. Classical
T Tauri stars in star forming regions were originally identified as strong
emission line objects in objective prism spectra. Later other techniques such as 
X-ray surveys could find similar, young stars but with much lower optical emission
line strengths than in classical TTS. As described in \citet{hart98} WTTS 
are not accreting from circumstellar disks.
In absence of this magnetospheric accretion, which would produce emission lines with
large velocity widths ($~>100$ \kms), solar-type magnetic energy dissipation
powers the excess (chromospheric) emission in WTTS so the observed lines are much
narrower and weaker than in the case of CTTS. \\

\begin{figure}
\includegraphics[angle=0,scale=0.8]{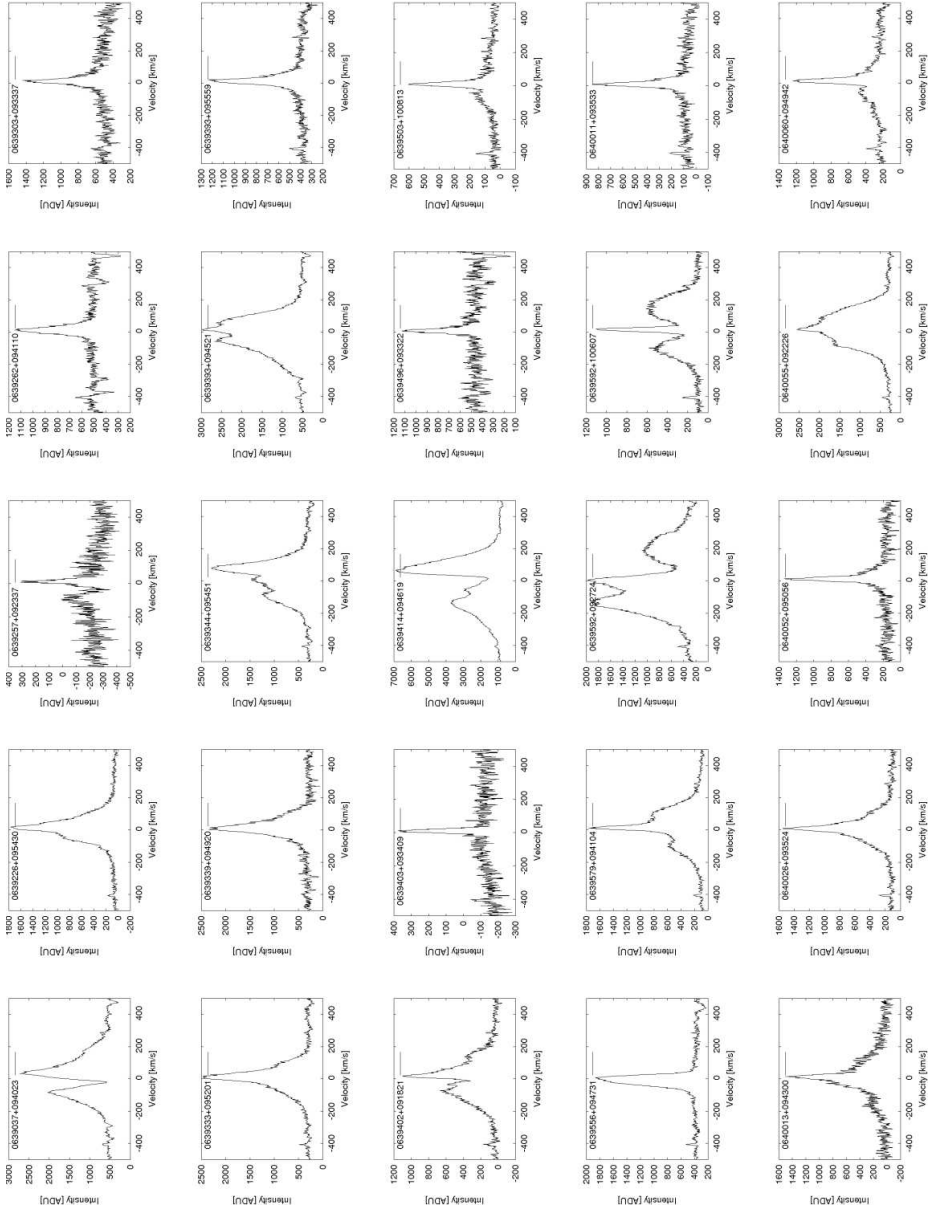}
\caption{Examples of $H\alpha$ profiles for stars with detected emission. Wavelength is converted to RV, and the 2MASS identification number is shown in each stamp.  The emission of the background is not subtracted, as we did not have separated offset sky exposures for all of our fields.
For all profiles see the electronic format of the paper or contact the authors. 
\label{fig:halp_p1}}
\end{figure}

We found 277 stars with $H\alpha$ emission exceeding the nebular component: 152 among
these were CTTS and 125 were WTTS, using the indicator of accretion from
\citet{whit03}, that the $H\alpha$ full width at 10\% intensity exceeded 270 \kms. 
A sample of $H\alpha$ profiles plotted on Fig. \ref{fig:halp_p1}.
We define membership as the RV value should be within 4 $\sigma$ of the cluster
mean velocity (22 \kms ~$\pm ~4\times$ 3.5 \kms). Based on this criteria
we found 178 of these emission stars to be radial velocity members (with an R value
larger than 2).
For these stars identification is given in Table \ref{table:hecto_targets_mem}.
A few of the H$\alpha$ stars listed in Table \ref{table:hecto_targets_mem}
have velocities close to the 4 $\sigma$ RV membership selection criteria, these
are noted as \textit{hsb}.
Further observations are needed to see if these are spectroscopic binaries.

Of the 277 obvious cluster member emission stars, 99 (35\%) have velocities outside
either $4\sigma$ lower or higher than the mean cluster velocity.
These stars yielded an R value less than 2, which suggest a random peak
in the cross-correlation happened to be close to the cluster velocity.
Based on their emission spectra these stars are young and therefore members,
but we do not have enough SNR in the cross-correlation to determine velocities.

We identified several spectroscopic binary stars. Some of these have resolved double peaks in the cross-correlation function, and we measured different velocities in the two separate observing runs. There are 6 such stars listed in Table
\ref{table:hecto_targets_mem} and \ref{table:hecto_targets_nonmem}
 with a three-letter note
\textit{``RDB''} (Resolved Double-lined Binary). In case of obvious side-lobes of the
cross-correlation function and two non-consistent but accurate velocities, we identified 
additional 3 stars as binaries (listed with a note \textit{``UDB''}, as Unresolved 
Double-lined Binary).
There are 6 \textit{``rdb''} and 5 \textit{``udb''} noted stars in Table \ref{table:hecto_targets_mem} and \ref{table:hecto_targets_nonmem},
for which we have just one RV measurement but observed double peaks or side-lobes in the
cross-correlation function.
11 stars showed no sign of binarity in the cross-correlation, but we got two significantly different velocities. These are likely single-lined binaries, and have an \textit{``sb''} note.

\subsection{Velocity correlations between stars and gas}

\begin{figure}
\includegraphics[angle=270,scale=0.5]{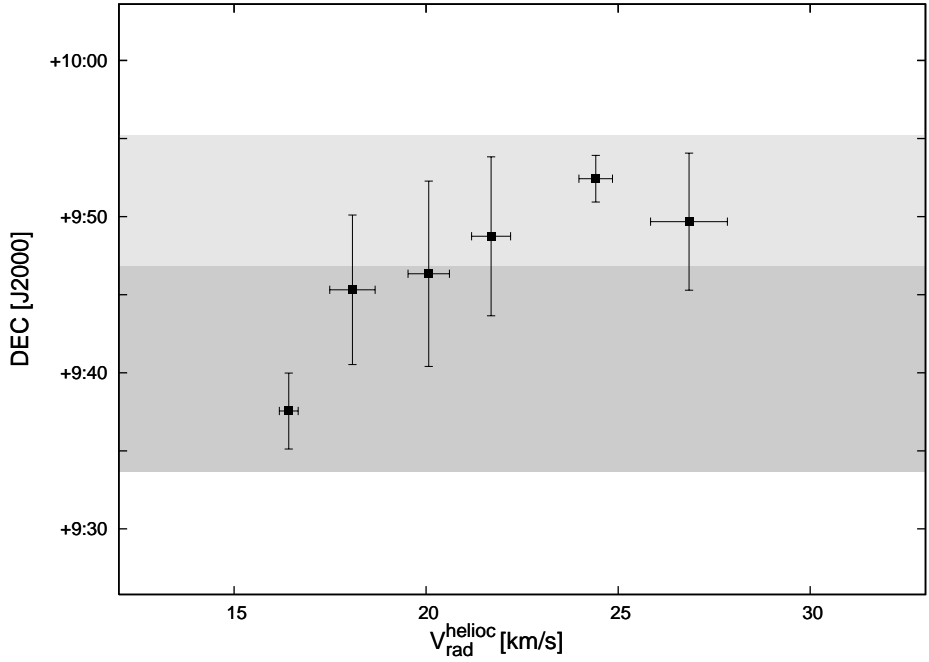}
\caption{A north-south velocity gradient observed in NGC 2264 is clearly shown,
by plotting the mean declination values in 2 \kms~ wide RV bins against the mean
radial velocity of the bins. Only stars observed at two epochs (and giving
consistent velocities) used for this plot, in order to decrease scatter. The
error bars represent the RMS of DEC/RV values of stars in a given velocity bin. The
small DEC error bar for the RV group of 24-26 \kms~ means that almost all the
stars in that velocity bin are at the northern-most part of the cluster.
The shaded areas show the declination ranges for the condensations of $H\alpha$ stars in R04.
\label{fig:radec}}
\end{figure}

The radial velocity distribution of stars in NGC 2264, as shown in Fig \ref{fig:hist},
is wide and asymmetric, probably due to subcluster groups with different
mean velocities. In optical images the cluster shows an elongated shape to the
north-south direction. The $H\alpha$ stars of R04 also show two denser
cores separated mainly to declination, one centered around 
$\alpha = 6^{h}41^{m}10^{s}$ and $\delta = +9^{\circ}30'$ and the other around
$\alpha = 6^{h}40^{m}50^{s}$ and $\delta = +9^{\circ}50'$ (see Fig. \ref{fig:optical}).
To explore this pattern further, in Fig. \ref{fig:radec}
we plot the declination values of stars observed at two epochs,
as a function of velocity.
In order to decrease uncertainties due to measurement errors, and natural
distribution in velocity space within a possible subcluster group, we
calculated mean declination values for 2 \kms~ wide bins in RV (Fig. \ref{fig:radec}).
The shaded areas show the declination ranges of the two largest (suggested)
subclusters based on the surface density of $H\alpha$ emission stars in the catalog
of R04. The southern-most declination limit of the shading was set by the
Hectochelle field observed in March.
The increasing velocity with increasing declination is obvious.

\begin{figure}
\includegraphics[angle=0,scale=0.8]{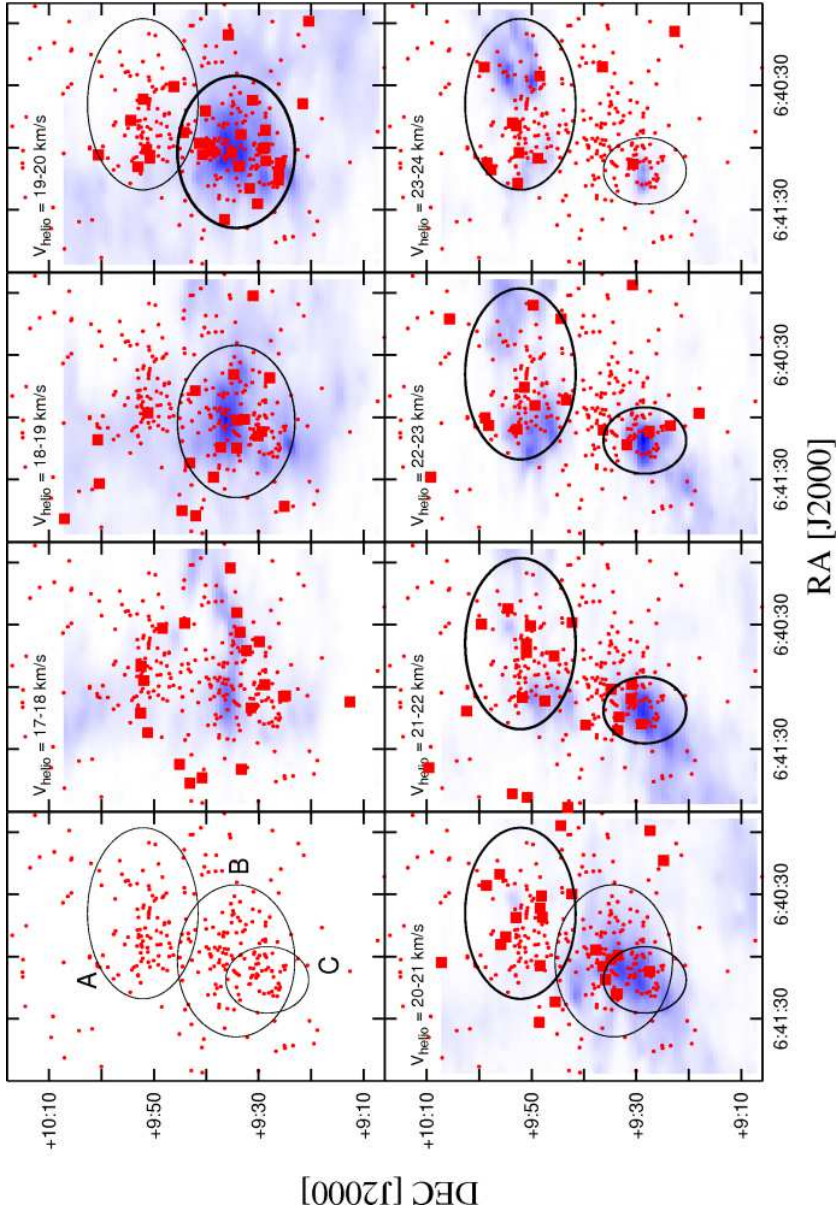}
\caption{ We find strong spatial correlation between the molecular
gas and stars in velocity space, by plotting radial velocity members of our sample (filled boxes)
superimposed on $^{13}$CO channel maps (both averaged in 1 \kms~ 
wide bins in RV). Positions of $H\alpha$ emission stars from R04 are also shown as dots to
draw an outline of the cluster (see upper left panel).
Three supposed subcluster structures (A,B,C) are also shown in the RV channels in which those are prominent.
\label{fig:13co_coord}}
\end{figure}

There is not a clear trend in RA, but this is consistent with the optical image of the cluster which shows a North-South elongation with only a very slight tilt in RA.
But the spatial pattern of stellar velocities becomes clear 
if we compare them to the pattern of molecular gas.
Using $^{13}$CO observations we tried to match the motion of the gaseous
component with the motion of the stellar component in NGC 2264. Fig. 
\ref{fig:13co_coord} displays channel maps as eight, 1 \kms~ wide panels centered
between $V_{LSR}=$ 3--10 \kms. Stars from the peak of Fig. \ref{fig:hist} 
with the same 1 \kms~ binning are superimposed on these maps; the mean
velocity of these groups are $V_{helioc}=$ 17--24 \kms.
The correlation between the position of dense $^{13}$CO cores and 
position of stars is very convincing.\\

On the upper-right panels of Fig. \ref{fig:13co_coord} there is a dense $^{13}$CO 
core at $\alpha = 6^{h}41^{m}00^{s}$ and $\delta = +9^{\circ}36'$. On the same panels, many stars occupy the same region (B) in the given RV bin. As we go to higher RV values, stars disappear from this area, as well as the gas, but two other dense $^{13}$CO regions appear. One is to the north, with an east-west elongation and a complicated structure, around the central point of $\alpha = 6^{h}40^{m}40^{s}$ and $\delta = +9^{\circ}52'$. Most of the stars fall in this region (A) on the lower panels of Fig. \ref{fig:13co_coord}, especially on the lower-right ones.\\

On the lower-middle panels of Fig. \ref{fig:13co_coord}, indicated by ``C'', there is
another dense $^{13}$CO 
core at $\alpha = 6^{h}41^{m}10^{s}$ and $\delta = +9^{\circ}29'$ at these higher velocity values.
This surrounds the Cone-nebula. On these panels, there are several stars very close to this region,
well separated in space from the larger number of stars to the north but lieing
in the same radial velocity range. These stars are responsible for the turn-down
of the north-south gradient on Fig. \ref{fig:radec}. This can
be a hint of a third subcluster group around the Cone, having a mean heliocentric 
velocity of $\approx 22$ \kms.\\

The distribution of R04 $H\alpha$~ stars also show correlations with the densest cores of molecular gas. At $V_{LSR}=9$~\kms~ there is a small, 3' diameter ring of stars around $\alpha = 6^{h}41^{m}10^{s}$ and $\delta = +9^{\circ}29'$, which surrounds a strong concentration of molecular gas best seen on that panel. The ring-like distribution of RV measured member stars on the $V_{LSR}=4$~\kms~ panel could also be a sign of some high energy process which played role in the formation of the cluster.

Similar agreements in the structure of stellar and gaseous components can be seen on Fig. \ref{fig:side}, which is a projection in RA, displaying the velocities along the horizontal axis and DEC values along the vertical axis. At $V_{LSR}=7$~\kms~ and $\delta = +9^{\circ}53'$ there is a``hole'' in the gas, likely a projection of a bubble, surrounded by stars in a ring which has higher stellar density than the neighboring regions. This structure likely has a connection with the star HD 47839 at $\alpha = 6^{h}40^{m}58^{s}$ and $\delta = +9^{\circ}53'44"$\\

\section{Discussion}

\begin{figure}
\plotone{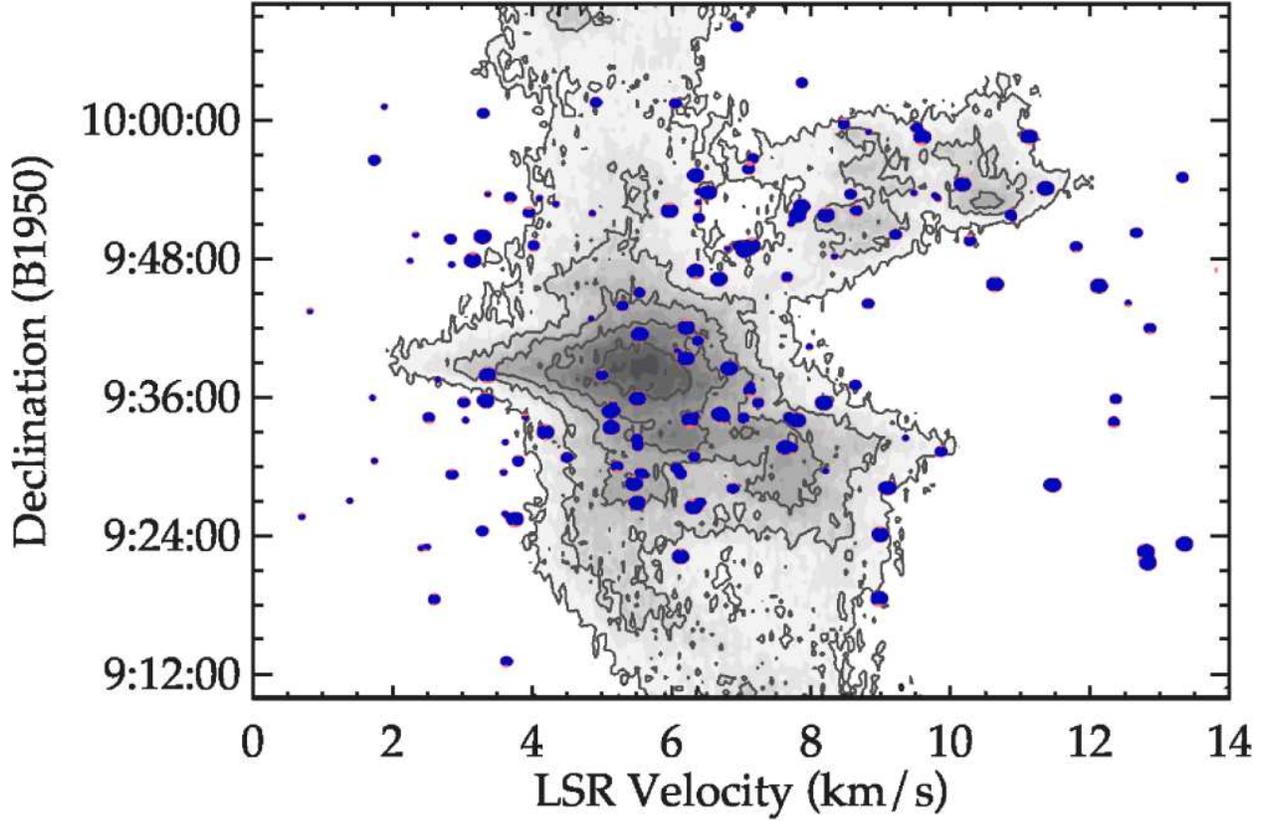}
\caption{ $^{13}$CO spatial-velocity plot of NGC 2264 projected along $\alpha$
shows an increased velocity dispersion at the northern end of the filament, a 
north-south gradient similar observed in the stellar component also plotted
on Fig. \ref{fig:radec}.
The molecular gas is shown as a shaded area, the dots are member stars with measured
velocities. The size of the dots correspond to the accuracy of RV measurement (bigger dot means higher accuracy). Note the bubble-like feature at V=7~\kms, $\delta = +9^{\circ}53'$, surrounded by stellar members. This structure likely connected to the bright (V=4.6) star HD 47839 at $\alpha = 6^{h}40^{m}58^{s}$ and $\delta = +9^{\circ}53'44"$
\label{fig:side}}
\end{figure}

While other star-forming regions have been shown to have significant substructure
in the young star population (e.g., Lada \& Lada 1995; Gomez \etal 1993), 
NGC 2264 is the first region for which we have sufficiently accurate measurements
to explore substructure in the stellar radial velocities.
The spatial distribution of stellar radial velocities and its correlation
with the velocity structure of the molecular gas shows that NGC 2264
is composed of relatively distinct components (Figures \ref{fig:13co_coord},\ref{fig:side}).  Intrinsic
errors due to various types of activity (starspots, accretion; Guenther et al. 2000)
are not large enough to obscure the strong correlation 
of the stellar velocities with the gas (Figures \ref{fig:13co_coord},\ref{fig:side}).
Though individual stars could have velocity offsets on the order of few \kms~ due to fast rotation and starspots/accretion \citep{neuh98,stra05}, that is very extreme and the large number of stars in our sample overrides the significance of such outliers.
 
The presence of such substructure in NGC 2264 is not 
surprising given its size and youth.  Assuming a local, within-group
velocity dispersion (one-dimensional) of order 2 \kms, 
and a spatial extent of $\pm$ 4 pc ($\sim \pm 17$~arcmin)
or so from the central region, the 
resulting crossing time of $\sim 2$~Myr is comparable to the estimated age of
the cluster (see Introduction).  Even adopting a velocity dispersion
of 3.5 \kms~ results in a crossing time of $\sim$ 1 Myr, close to the youngest
stellar age estimates.  Numerical simulations show \citep{scal02}
that substructure in a stellar cluster can remain detectable even after
several crossing times, depending upon the number of stars involved and
their initial subcluster densities.  However, it is unlikely that
the region is considerably older than one or two crossing times,
because the gas also shows the same substructure as the stars.
The cluster gas will shock and dissipate considerable amounts of energy on the
first crossing, and thus the initial substructure should be damped out much
more rapidly than for the stellar population. 

\citet{crut78} presented observational evidence
for two separate gaseous cores inside NGC 2264
which have a mean velocity difference of 2~\kms. These two cores
correspond in position with the centers of our southern subclusters 
\textit{B} and \textit{C} (\textit{C} is in the
direct vicinity of the Cone nebula), see lower left panel on Fig. \ref{fig:13co_coord}.
The velocity gradient we estimate based on our observations
agrees in direction and in its amount with that
determined by \citet{crut78}. \citet{lang80} verified the same gradient
based on NH$_3$ measurements and also explained the motion of the two cores
as rotation, with a period of 4 Myrs if the rotational 
axis is perpendicular to the line of sight.  However, as the
$^{13}$CO observations of \citet{ridg03} show (Fig. \ref{fig:13co_coord}), the
velocity field is more complicated than simply that of pure rotation.

Large velocity dispersions sometimes arise in molecular gas as the
result of powerful bipolar outflows or other stellar energy input.
However, the systematic motions seen in the gas require too much
energy to explain in this way.  Instead, we interpret the velocities
as mostly gravitationally-generated, which would very simply explain why the
stars spatially co-located with the gas show the same velocities.

\begin{figure}
\plotone{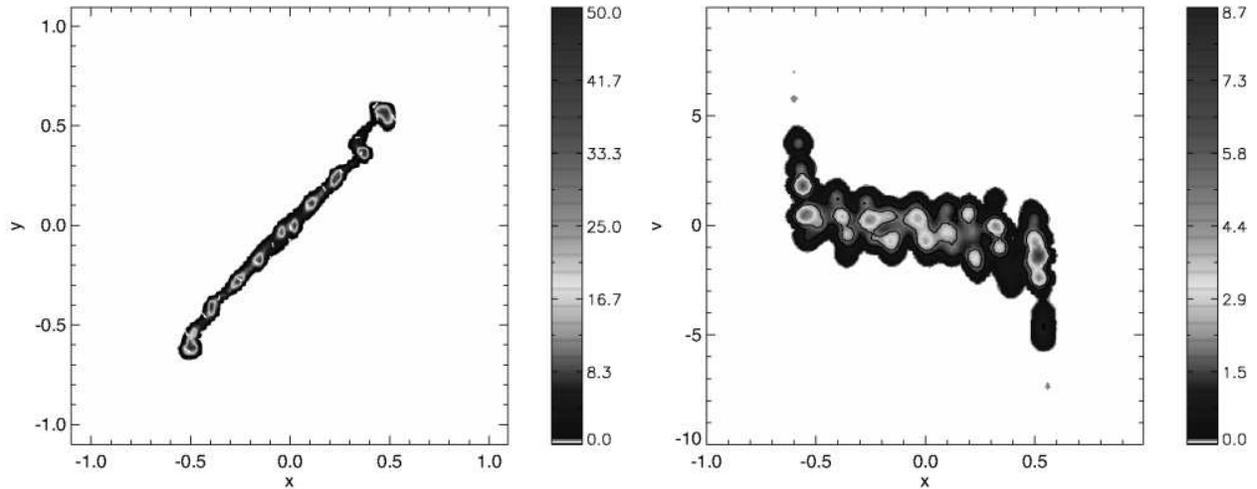}
\caption{ Simulation of gaseous sheet collapse to a filament
(see Burkert \& Hartmann 2004). The left panel shows the spatial
distribution of the filament, which exhibits a number of condensations which
resulted from slight substructure in the initial conditions, subsequently
magnified by gravitational collapse from an initially elliptical configuration.
The right panel shows the projected emission along one axis (assumed 
optically-thin) as a function of radial velocity.  The various condensations
show offset velocities which are the result of gravitational collapse from
opposite sides of the filament. The scaling to the mass and radius of the 
NGC 2264 gaseous filament is such that each unit of velocity corresponds to
1.6 \kms.  The existence of condensations with differing velocities at the
$\sim 2$ \kms level, plus the higher-velocity tails at the end of the
filament - the result of gravitational attraction on at "focal points"
(see Burkert \& Hartmann 2004) where especially large density concentrations
develop - is suggestive of the structure seen in both the gas and stars of 
NGC 2264 (Fig. \ref{fig:side}) (see text).
\label{fig:simul}}
\end{figure}

\citet{bh04} recently explored a simple model of the formation of
star-forming filaments.  Their calculations assumed that molecular clouds
were formed schematically from colliding supersonic flows which formed (approximately)
sheet-like configurations.  If the sheet is elongated, the result is
gravitational collapse first along the shortest dimension, forming
a filamentary structure.  As shown in the left panel of 
Fig. \ref{fig:simul} (reproduced from Figure 10 
in Burkert \& Hartmann 2004), irregularities initially present in the sheet 
grow to form non-linear density enhancements by the time the configuration 
collapses to a filament.  The right panel of Fig \ref{fig:simul} shows the integrated
surface density as a function of the radial velocity in the x-direction,
showing that multiple dense structures with coherent velocities are present.
Often individual ``clumps'' appear in nearby pairs with distinct velocities.
This pattern is the result of the overall collapse of the elongated sheet;
material originating on opposite sides of the sheet fall in with opposing
velocities.  The high-velocity infall at each end of the filament is due
to the highest gravitational acceleration at filament ends \citep{bh04}, 
coupled with some initial rotation of the sheet.

We suggest that the Burkert \& Hartmann model qualitatively explains the
velocity structure seen in the NGC 2264 region.  The pattern of first
higher velocities, then lower velocities, then somewhat higher again
proceeding southward along the region (Figure 7) can be explained by
gravitationally-driven infall from opposite sides of a cloud falling
in to form a more filamentary structure.  The highest velocity structure
seen at the northern end may be similar to the acceleration seen in the
end of the filament in the right panel of Fig. \ref{fig:simul}.  The velocity units
in Figure 7 are $(G M R/\pi)^{1/2}$; if we adopt a total gas mass of
$4000 M_{\sun}$ from \citet{ridg03} and a filament half-length of 4 pc,
these units correspond to $\sim 2~$\kms.  Clearly, gravitational collapse
can account for the magnitude of the motions seen in NGC 2264,
especially considering geometric uncertainties and the simplicity of the
model.

\section{Summary}

We have carried out a  
spectroscopic study of stars in NGC 2264, utilizing the Hectochelle multiobject
spectrograph on the 6.5 m MMT. Obtaining 1078 spectra of 990 stars we classified 471 stars as cluster members based on their radial velocity and/or $H\alpha$~ emission. We determined radial velocities with an accuracy of $<1.5$~\kms~ for 344 members. 
The radial velocity distribution of these stars is wide, non-gaussian,
and the dispersion is 3.5~\kms.  We compared the stellar radial velocities
to the velocity of the molecular gas in the cluster, traced by $^{13}$CO,
and found a strong correlation in the spatial distribution of the two components.
At least three subgroups of stars can be identified with distinct
spatial and velocity coherence.  We interpret this substructure as
the result of gravitational collapse of initial clumps of star-forming
gas from a more extended structure to a roughly filamentary distribution. 
Using the results presented here, more advanced numerical simulations could
help elucidate the initial conditions that produced NGC 2264.

\acknowledgments
This work was supported in part by NASA grant NAG5-13210.

\clearpage

\clearpage

\clearpage

\clearpage

\clearpage

\clearpage

\clearpage

\clearpage

\begin{deluxetable}{ccccc}
\tabletypesize{\scriptsize}
\rotate
\tablecaption{Summary of spectroscopic observations \label{table:obs}}
\tablewidth{0pt}
\tablehead{
\colhead{Date} & \colhead{field center} & \colhead{exposures}
& \colhead{offset sky} & \colhead{number of targets}\\ 
}
\startdata
2004 Mar 10	&    $6^{h}~40^{m}~51^{s}~~ +09\degr~44\arcmin~46\arcsec$ &  $4\times15$ min  &  $1\times15$ min  & 155\\
2005 Dec 02	&    $6^{h}~40^{m}~46^{s}~~ +09\degr~38\arcmin~13\arcsec$ &  $3\times15$ min  &  ---  & 234 \\
2005 Dec 02	&    $6^{h}~40^{m}~35^{s}~~ +09\degr~36\arcmin~56\arcsec$ &  $3\times15$ min  &  ---  & 232\\
2005 Dec 02	&    $6^{h}~40^{m}~42^{s}~~ +09\degr~45\arcmin~36\arcsec$ &  $3\times15$ min  &  ---  & 231\\
2005 Dec 03	&    $6^{h}~40^{m}~42^{s}~~ +09\degr~32\arcmin~13\arcsec$ &  $3\times15$ min  &  ---  & 226\\
\enddata
\end{deluxetable}

\begin{deluxetable}{cccrrrr}
\tabletypesize{\scriptsize}
\rotate
\tablecaption{Observed radial velocity standard stars
in a sub-field of Selected Area 57 \label{sa57}}
\tablewidth{0pt}
\tablehead{
\colhead{Name} & \colhead{RA} & \colhead{DEC} & \colhead{$V_{rad}$} 
& \colhead{$V_{rot}$} & \colhead{$T_{eff}$} & \colhead{log(g)}\\
}
\startdata

W33245	&  13:03:50.9  &  +30:11:38  &  -20.95  &  0.0  &  5000 & 3.5  \\
W22942  &  13:04:25.2  &  +30:25:44  &  -16.32  &  3.3  &  4500 & 2.5  \\
W24128  &  13:04:45.0  &  +29:51:22  &  2.20  &  3.2  &  5500 & 4.5  \\
W23082  &  13:04:58.4  &  +30:10:39  &  -21.68  &  3.5  &  6250 & 4.5  \\
W23108  &  13:05:18.4  &  +30:10:46  &  -16.43  &  7.0  &  4250 & 2.0  \\
W24226  &  13:05:23.5  &  +29:43:02  &  11.69  &  4.1  &  6500 & 4.5  \\
W23131  &  13:05:29.3  &  +30:11:12  &  -5.96  &  5.4  &  5500 & 3.5  \\
W23961  &  13:06:00.6  &  +30:01:22  &  -6.98  &  4.2  &  6000 & 4.5  \\
W23870  &  13:06:45.8  &  +29:46:30  &  -14.85  &  5.3  &  4750 & 4.0  \\
W23833  &  13:06:58.6  &  +29:57:49  &  -4.90  &  0.5  &  5250 & 4.0  \\

\enddata
\tablecomments{Table \label{table:sa57}
\textit{$V_{rad}$} --- heliocentric
velocities on the native CfA velocity system.  To convert to an
absolute velocity system based on extensive observations of minor
planets, 0.139 \kms\ should be added to the native CfA
velocities \citep{stef99}.}
\end{deluxetable}

\begin{deluxetable}{crrrrrrcccl}
\tabletypesize{\scriptsize}
\rotate
\tablecaption{Members \label{table:hecto_targets_mem}}
\tablewidth{0pt}
\tablehead{
\colhead{2MASS\_id} & \colhead{REBU} & \colhead{REIP} & \colhead{LAMM} 
& \colhead{$J$} & \colhead{$(J-H)$} & \colhead{$(H-K)$} & \colhead{$H\alpha$} &
\colhead{$RV_{Dec}/RV_{Mar}$} & \colhead{$R_{Dec}/R_{Mar}$} & \colhead{notes} \\
}
\startdata
0639037+094023	& --	& 325	& --	& 12.65	& 0.74	& 0.35	& c	& 20.46/--	& 7.6/--	& H\\
0639156+093912	& --	& --	& --	& 13.26	& 0.44	& 0.11	& --	& 17.16/--	& 4.0/--	& H\\
0639157+092929	& --	& --	& --	& 11.67	& 0.69	& 0.24	& --	& 33.80/--	& 12.3/--	& H\\
0639170+095815	& --	& --	& --	& 12.43	& 0.36	& 0.17	& --	& 14.06/--	& 6.2/--	& H\\
0639226+091703	& --	& --	& --	& 11.13	& 0.19	& 0.14	& --	& 28.28/--	& 2.2/--	& H\\
0639226+095430	& --	& 330	& --	& 14.25	& 0.61	& 0.35	& w	& 16.51/--	& 2.1/--	& H\\
0639229+093616	& --	& --	& --	& 12.62	& 0.49	& 0.14	& --	& 14.49/--	& 11.3/--	& H\\
0639246+092233	& --	& --	& --	& 12.79	& 0.41	& 0.12	& --	& 18.30/--	& 9.7/--	& H\\
0639262+094110	& R453	& --	& --	& 12.71	& 0.58	& 0.23	& w	& 16.17/--	& 6.2/--	& H\\
0639266+093037	& --	& --	& --	& 15.50	& 0.66	& 0.25	& --	& 29.76/--	& 2.5/--	& H\\
0639275+092309	& --	& --	& --	& 11.03	& 0.91	& 0.31	& --	& 34.54/--	& 11.2/--	& H\\
0639278+092839	& --	& --	& --	& 10.81	& 0.54	& 0.20	& --	& 13.17/--	& 11.6/--	& H\\
0639293+092058	& --	& --	& --	& 12.20	& 0.41	& 0.12	& --	& 25.38/--	& 11.2/--	& H\\
0639303+093337	& --	& --	& --	& 12.93	& 0.69	& 0.14	& w	& 21.27/--	& 3.6/--	& H\\
0639329+095629	& R691	& --	& --	& 13.16	& 0.46	& 0.13	& --	& 23.95/23.68	& 10.7/9.4	& H\\
0639333+095201	& R705	& --	& --	& 12.63	& 0.73	& 0.24	& c	& 19.32/--	& 4.1/--	& H\\
0639335+095140	& R712	& --	& --	& 13.71	& 0.30	& 0.30	& --	& 25.61/25.83	& 3.1/3.0	& H\\
0639339+094920	& R729	& --	& --	& 13.04	& 0.77	& 0.17	& w	& 19.92/--	& 5.3/--	& H\\
0639344+095451	& --	& --	& --	& 12.99	& 0.78	& 0.28	& c	& 19.25/--	& 6.4/--	& H\\
0639347+094654	& --	& --	& --	& 14.45	& 0.59	& 0.26	& --	& 13.30/--	& 2.8/--	& H\\
0639353+093232	& --	& --	& --	& 13.21	& 0.44	& 0.14	& --	& 28.06/--	& 2.9/--	& vm\\
0639360+092426	& --	& --	& --	& 11.85	& 0.67	& 0.15	& --	& 13.90/--	& 11.6/--	& H\\
0639388+095151	& R893	& --	& --	& 12.47	& 0.40	& 0.23	& --	& 13.61/--	& 3.0/--	& H\\
0639393+094521	& R908	& --	& --	& 12.90	& 0.86	& 0.50	& c	& 19.26/--	& 7.8/--	& H\\
0639393+095559	& R905	& --	& --	& 12.79	& 0.65	& 0.24	& w	& 18.32/--	& 4.2/--	& --\\
0639414+094619	& R985	& --	& --	& 12.57	& 0.65	& 0.22	& c	& 18.95/--	& 7.5/--	& H\\
\enddata
\tablecomments{Hectochelle targets in NGC 2264 found to be members based on measured RV value
or detected H$\alpha$ emission. The criteria of beeing RV member is to have at least one
velocity measurement within 4$\sigma$ of the cluster mean velocity:
8 \kms~ $< ~V_{helio} <$ 36 \kms~. (Stars listed only with R value of cross correlation larger than 2.)
For the entire table see the electronic version of the Jurnal, or contact the authors.\\
\textbf{2MASS id} --- identification number from 2MASS catalog
(truncated RA and DEC coordiantes as: HHMMSSS+DDMMSS);
\textbf{REBU} --- id numbers from Rebull et al 2002, \aj 123, 1528; 
\textbf{REIP} --- id from Reipurth et al 2004 \aj 127, 1117;
\textbf{LAMM} --- id from Lamm et al 2004 A\&A 417, 557;
\textbf{J} --- 2MASS J magnitude and;
\textbf{(J-H)} --- 2MASS $(J-H)$ color index; 
\textbf{(J-H)} --- 2MASS $(H-K)$ color index;\\
\textbf{H$\alpha$} ---  $H\alpha$~ emission detected: \textbf{c} -- CTTS; \textbf{w} -- WTTS; 
\textbf{x} -- V642 Mon; \textbf{r} -- reduction error in 2004 Dec data set, categorized as WTTS in 2004 Mar data;\\
\textbf{VR$_{Dec}$/VR$_{Mar}$} --- heliocentric radial velocity measured 2004 Dec / 2004 Mar;
\textbf{R$_{Dec}$/R$_{Mar}$} --- R value of cross correlation (see text for details) 2004 Dec / 2004 Mar;\\
\textbf{notes} --- MEMBERSHIP: 
\textbf{vm} -- at least one RV measurement suggest membership, but R value is low;
\textbf{hm} -- just $H\alpha$\ emission suggest membership, none of the RV values (if there are two);
ACCURACY: \textbf{H} -- most of the templates gave the same velocity, R value of cross-correlation is high, accurate RV value (typical error $< 1.5$~\kms) for one RV measurement;
BINARITY: \textbf{RDB} -- resolved, double lined binary, with RV at 2 epochs (separated double peak in correlation function);
\textbf{UDB} -- unresolved double lin. binary, with RV at 2 epochs (blended/side lobed peak, but clear asymmetry in correlation function);
\textbf{rdb} -- same as for RDB, but only 1 RV measurement;
\textbf{udb} -- same as for UDB, but only 1 VRV measurement;
\textbf{sb} -- likely single lined binary, RV at two epochs and at least one suggest membership, some also have $H\alpha$~ emission;
\textbf{hsb} -- member based on $H\alpha$ emission detected, but correlation either had
just acceptable R value (barely larger than 2) or radial velocity is close to the edge of RV distribution.
}
\end{deluxetable}

\begin{deluxetable}{crrrrrrcccl}
\tabletypesize{\scriptsize}
\rotate
\tablecaption{Non-members \label{table:hecto_targets_nonmem}}
\tablewidth{0pt}
\tablehead{
\colhead{2MASS\_id} & \colhead{REBU} & \colhead{REIP} & \colhead{LAMM} 
& \colhead{$J$} & \colhead{$(J-H)$} & \colhead{$(H-K)$} & \colhead{$H\alpha$} &
\colhead{$RV_{Dec}/RV_{Mar}$} & \colhead{$R_{Dec}/R_{Mar}$} & \colhead{notes} \\
}
\startdata
0639152+093937	& R3	& --	& --	& 10.90	& 0.76	& 0.27	& --	& 45.53/--	& 5.5/--	& H\\
0639161+092105	& --	& --	& --	& 9.95	& 1.00	& 0.36	& --	& 85.84/--	& 11.0/--	& --\\
0639163+091953	& --	& --	& --	& 12.61	& 0.29	& 0.15	& --	& 65.74/--	& 2.8/--	& --\\
0639174+092137	& --	& --	& --	& 13.61	& 0.32	& 0.18	& --	& 44.93/--	& 2.3/--	& H\\
0639183+093336	& --	& --	& --	& 14.88	& 0.71	& 0.13	& --	& 42.58/--	& 2.7/--	& H\\
0639200+093036	& --	& --	& --	& 11.32	& 0.70	& 0.24	& --	& 49.10/--	& 5.2/--	& H\\
0639229+093523	& --	& --	& --	& 12.41	& 0.45	& 0.13	& --	& -12.36/--	& 9.0/--	& H\\
0639239+094201	& R353	& --	& --	& 10.84	& 1.09	& 0.44	& --	& 81.66/--	& 11.3/--	& --\\
0639240+092550	& --	& --	& --	& 11.77	& 0.67	& 0.19	& --	& 44.22/--	& 5.4/--	& H\\
0639241+091805	& --	& --	& --	& 13.48	& 0.37	& 0.15	& --	& -19.37/--	& 5.0/--	& H\\
0639243+095619	& --	& --	& --	& 10.68	& 0.74	& 0.28	& --	& -12.91/--	& 11.6/--	& H\\
0639245+095040	& --	& --	& --	& 10.63	& 0.26	& 0.07	& --	& 85.41/--	& 10.4/--	& --\\
0639246+091721	& --	& --	& --	& 12.84	& 0.81	& 0.34	& --	& 69.63/--	& 6.2/--	& --\\
0639250+095339	& --	& --	& --	& 13.06	& 0.51	& 0.14	& --	& 78.26/--	& 3.9/--	& rdb\\
0639275+095100	& --	& --	& --	& 11.67	& 0.79	& 0.28	& --	& 87.58/--	& 12.1/--	& --\\
0639283+095358	& --	& --	& --	& 10.06	& 0.82	& 0.31	& --	& 0.78/--	& 7.5/--	& H\\
0639287+092640	& --	& --	& --	& 12.40	& 0.28	& 0.16	& --	& -2.97/--	& 2.5/--	& H\\
0639303+092300	& --	& --	& --	& 12.05	& 0.47	& 0.10	& --	& 109.22/--	& 10.9/--	& --\\
0639304+094924	& --	& --	& --	& 10.03	& 0.77	& 0.25	& --	& -9.34/--	& 12.4/--	& H\\
0639311+095330	& --	& --	& --	& 10.32	& 0.28	& 0.08	& --	& 4.17/--	& 9.8/--	& H\\
0639333+093011	& --	& --	& --	& 13.56	& 0.47	& 0.17	& --	& 64.57/--	& 4.1/--	& --\\
0639334+094728	& R707	& --	& --	& 13.36	& 0.40	& 0.20	& --	& 37.26/--	& 2.0/--	& H\\
0639344+092338	& --	& --	& --	& 12.13	& 0.38	& 0.10	& --	& 99.31/--	& 4.8/--	& --\\
0639346+092440	& --	& --	& --	& 12.72	& 0.35	& 0.07	& --	& 52.81/--	& 5.5/--	& --\\
0639373+092637	& --	& --	& --	& 11.47	& 0.63	& 0.17	& --	& 39.21/--	& 10.2/--	& H\\
0639386+094044	& --	& --	& --	& 12.74	& 1.27	& 0.57	& --	& 52.73/--	& 2.0/--	& --\\
0639396+094544	& R916	& --	& --	& 11.98	& 0.62	& 0.19	& --	& --/43.05	& --/7.2	& --\\
\enddata
\tablecomments{Hectochelle targets in NGC 2264 found to be non
members based on RV measurement. (Stars listed only with R value of cross correlation larger than 2.)
For the entire table see the electronic version of the Jurnal, or contact the authors.\\
\textbf{2MASS id} --- identification number from 2MASS catalog
(truncated RA and DEC coordiantes as: HHMMSSS+DDMMSS);
\textbf{REBU} --- id numbers from Rebull et al 2002, \aj 123, 1528; 
\textbf{REIP} --- id from Reipurth et al 2004 \aj 127, 1117;
\textbf{LAMM} --- id from Lamm et al 2004 A\&A 417, 557;
\textbf{J} --- 2MASS J magnitude and;
\textbf{(J-H)} --- 2MASS $(J-H)$ color index; 
\textbf{(J-H)} --- 2MASS $(H-K)$ color index;\\
\textbf{H$\alpha$} ---  $H\alpha$~ emission detected: \textbf{c} -- CTTS; \textbf{w} -- WTTS; 
\textbf{x} -- V642 Mon; \textbf{r} -- reduction error in 2004 Dec data set, categorized as WTTS in 2004 Mar data;\\
\textbf{VR$_{Dec}$/VR$_{Mar}$} --- heliocentric radial velocity measured 2004 Dec / 2004 Mar;
\textbf{R$_{Dec}$/R$_{Mar}$} --- R value of cross correlation (see text for details) 2004 Dec / 2004 Mar;\\
ACCURACY: \textbf{H} -- most of the templates gave the same velocity, R value of cross-correlation is high, accurate RV value (typical error $< 1.5$~\kms) for one RV measurement;
BINARITY: \textbf{RDB} -- resolved, double lined binary, with RV at 2 epochs (separated double peak in correlation function);
\textbf{UDB} -- unresolved double lin. binary, with RV at 2 epochs (blended/side lobed peak, but clear asymmetry in correlation function);
\textbf{rdb} -- same as for RDB, but only 1 RV measurement;
\textbf{udb} -- same as for UDB, but only 1 VRV measurement;
\textbf{sb} -- likely single lined binary, RV at two epochs and at least one suggest membership, some also have $H\alpha$~ emission;
\textbf{hsb} -- member based on $H\alpha$ emission detected, but have unusable RV measurement due to low SNR; only three of these have trustable velocities with $R>2.5$ (see text).
}
\end{deluxetable}


\begin{thebibliography}{}

\bibitem[Bate et al.(1998)]{bate98}
Bate, M.R., Clarke, C.J., \& McCaughrean, M.J.
 1998, \mnras, 297, 1163

\bibitem[Bate et al.(2003)]{bate03} 
Bate, M.R., Bonnell, I.A., \& Bromm, V.,
 2003, \mnras, 339, 577 

\bibitem[Bessel \& Brett(1988)]{bess88}
Bessel, M.S., \& Brett, J.M.
 1988. \pasp, 100, 1134

\bibitem[Burkert \& Hartmann(2004)]{bh04} Burkert, A., \& 
Hartmann, L.
 2004, \apj, 616, 288

\bibitem[Crutcher \& Hartkopf(1978)]{crut78}
Crutcher, R.M., \& Hartkopf, W.I.
 1978, \apj, 226, 839 
 
\bibitem[Fabricant et al.(2004)]{fabr04}
Fabricant, D.G., et al.
 2004, \procspie, 5492, 767

\bibitem[Fabricant et al.(2006)]{fabr06}
Fabricant, D.G., et al.
 2006, \pasp, In Press

\bibitem[Flaccomio et al.(2000)]{flac00}
Flaccomio, E., Micela, G., Sciortino, S., Damiani, F., Favata, F.,
Harnden, F.R. Jr., \& Schachter, J.
 2000, \aap, 355, 651
 
\bibitem[Gomez et al.(1993)]{gome93}
Gomez, M., Hartmann, L., Kenyon, S., \& Hewtett, R.,
 1993, \aj, 105, 1927 

\bibitem[Guenther et al.(2000)]{guen00}
Guenther, E.W., Joergens, V., Neuh\"{a}user, R., Torres, G., Batalha, N.S.,
Vijapurkar, J., Fern\'{a}ndez, M., \& Mundt, R., 2001, in
Zinnecker, H., \& Mathieu, R.D., eds, Iau Symp. 200, The Formation of Binary
Stars. ASP, Provo, p. 165
  
\bibitem[Harris(1976)]{hari76} 
Harris, G.L.H
 1976, \apjs, 30, 451
 
\bibitem[Hartmann(1998)]{hart98} 
Hartmann, L.
 1998, Accretion Process in Star Formation, Cambridge University Press

\bibitem[Hartmann(2003)]{hart03} Hartmann, L.\ 2003, \apj, 
585, 398

\bibitem[Herbig(1954)]{herb54}
Herbig, G.H.
 1954, \apj, 119, 483
 

\bibitem[Hillenbrand et al.(1998)]{hill98}
Hillenbrand, L.A., Strom, S., Calvet, N., Mermill, K.M., Gatley, I., Makidon, R.,
Meyer, M., \& Skrutskie, M.
 1998, \aj, 116, 1816

\bibitem[Joergens \& Guenther(2001)]{joer01} 
Joergens, V., \& Guenther, E.
 2001, \aap, 379, L9  
 
 
\bibitem[Jones \& Walker(1988)]{jone88} 
Jones, B.F., \& Walker, M.F.
 1988, \aj, 95, 6 

\bibitem[Kurucz(1979)]{kuru79}
Kurucz, R., 1979, ApJS, 40, 1
 
\bibitem[Kurtz \& Mink(1998)]{kurt98}
Kurtz, M.J., \& Mink, D.J.
 1998, \pasp, 110, 934

\bibitem[Lada \& Lada(1995)]{lada95}
Lada, E.A., \& Lada, C.J.
 1995, \aj, 109, 1682  
 
\bibitem[Lada et al.(1993)]{lada93}
Lada, C., Young, E., \& Greene, T.
 1993, \apj, 408, 471 

\bibitem[Lamm et al.(2004)]{lamm04} 
Lamm, M.H., Bailer-Jones, C.A.L., Mundt, R., Herbst, W., \& Scholz, A.
 2004, \aa, 417, 557

\bibitem[Lang \& Willson(1980)]{lang80}
Lang, K.R., \& Willson, R.F.
 1980, \apj, 238, 867
  
\bibitem[Latham(1992)]{Lath92}
Latham, D.W. 
1992, in ASP Conf. Ser. 32, IAU Coll. No. 135, Complementary Approaches 
to Binary and Multiple Star Research, eds. H. McAlister \& W. Hartkopf, 
p. 110
 
\bibitem[Latham et al.(1991)]{Latetal91}
Latham, D.W., Davis, R.J., Stefanik, R.P., Mazeh, T. \& Abt, H.A.
1991, \aj, 101, 625
   
\bibitem[Meyer et al.(1997)]{meye97}
Meyer, M.R., Calvet, N., \& Hillenbrand, L.H.
 1997, \aj, 114, 288 

\bibitem[Muzerolle et al.(2003)]{muze03}
Muzerolle, J., Calvet, N., Hartmann, L., \& D'Alessio, P.
 2003, \apj, 597, 149  

\bibitem[Neuhaeuser et al.(1998)]{neuh98} 
Neuhaeuser, R., Wolk, S.J., Torres, G., Preibisch, Th., Stout-Batalha, N.M.,
Hatzes, A.P., Frink, S., Wichmann, R., Covino, E., Alcala, J.M., Brandner, W., Walter, F.M., Sterzik, M.F., \& Koehler, R.
  1998, A\&A, 334, 873
 
\bibitem[Ogura(1984)]{ogur84}
Ogura, K.
 1984, \pasj, 36, $1390$
   

\bibitem[Palla \& Stahler(1999)]{palla99} 
Palla, F. \& Stahler, S.W. 1999, \apj, 525, 772


\bibitem[Park et al.(2000)]{park00} 
Park, B.-G., Sung, H., Bessel, M.S., \& Kang, Y.H
 2000, \aj, 120, 894

\bibitem[P\'{e}rez et al.(1987)]{pere87}
P\'{e}rez, M.R., Th\'{e}, P.S., \& Westerlund, B.E.
 1987, \pasp, 99, 1050
 
\bibitem[Pickering and Kapteyn(1918)]{pick18}
Pickering. E.C., \& Kapteyn. J.C.
 1018, Annals of Harvard College Observatory, vol. 101, pp.1-368
  
\bibitem[Ram\'irez et al.(2004)]{rami04} 
Ramírez, S.V., Rebull, L., Stauffer, J., Strom, S., Hillenbrand, L., 
Hearty, T., Kopan, E.L., Pravdo, S., Makidon, R., \& Jones, B.
 2004, \aj, 128, 787
 
\bibitem[Reipurth et al.(2004)]{reip04} 
Reipurth, B., Pettersson, B., Armond, T., Bally, J., \& Vaz, L.P.R.,
2004, \aj, 127, 1117 
 
\bibitem[Rebull et al.(2002)]{rebu02}
Rebull, L.M., Makidon, R.B., Strom, S.E., Hillenbrand, L.A., Birmingham, A.,
Patten, B.M., Jones, B.F., Yagi, H., \& Adams, M.T.
 2002, \aj, 123, 1528

\bibitem[Ridge et al.(2003)]{ridg03}
Ridge, N.A., Wilson, T.L., Megeath, S.T., Allen, L.E., \& Myers, P.C.
 2003, \aj, 126, 286

\bibitem[Sagar \& Joshi(1983)]{saga83} 
Sagar, R., \& Joshi, U.C.
 1983, \mnras, 205, 747

\bibitem[Scally \& Clarke(2002)]{scal02}
Scally, A., \& Calrke, C.
 2002, \mnras, 334, 156 
  
\bibitem[Sicilia-Aguilar et al.(2004)]{auro04}
Sicilia-Aguilar, S., Hartmann, L.W., Szentgyorgyi, A.H., Fabricant, D.G.,
F\H{u}r\'{e}sz, G., Roll, J.B., Conroy, M.A., Calvet, N., Tokarz, S.,
\& Hern\'andez, J.
 2005, \aj, 129, 363

\bibitem[Sung et al.(1997)]{sung97}
Sung, H., Bessell, M.S., \& Lee, S.-W.
 1997, \aj, 114, 2644
 
\bibitem[Sung et al.(2004)]{sung04}
Sung, H., Bessell, M.S., \& Chun, M-Y.
 2004, \aj, 128, 1684
 
\bibitem[Stefanik, Latham, \& Torres(1999)]{stef99}
Stefanik, R. P., Latham, D. W. \& Torres, G.
1999, in ASP Conf. Ser. 185, IAU Coll. No. 170, Precise Stellar Radial 
Velocities, eds. J. B. Hearnshaw \& C. D. Scarfe, p. 354

\bibitem[Strassmeier et al.(2005)]{stra05}
Strassmeier, K.G., Rice, J.B., Ritter, A., K\"{u}ker, M., Hussain, G.A.J., Hubrig, S., \& Shobbrook, R.
 2005, A\&A, 440, 1105

\bibitem[Szentgyorgyi et al.(1998)]{sain98}
Szentgyorgyi, A.H., Cheimets, P., Eng, R., Fabricant, D.G., Geary, J.C.,
Hartmann, L., Pieri, M.R., \& Roll, J.B. 
 1998, \procspie, 3355, 242 

\bibitem[Tonry \& Davis(1979)]{tonr79}
Tonry, J. \& Davis, M.
 1979, \aj, 84, 1511

\bibitem[Vasilevskis et al.(1965)]{vasi65} 
Vasilevskis, S., Sanders, W.L., \& Balz, A.G.A. Jr.
 1956, \aj, 70, 797
  
\bibitem[Walker(1956)]{walk56}
Walker, M.F.
 1956, \apjs, 2, 365 
 
\bibitem[White \& Basri(2003)]{whit03}
White, R.J., \& Basri, G.
 2003, \aj, 582, 1109
 
\end{thebibliography}
\end{document}